\documentclass{article}
\usepackage[a4paper, total={5.9in, 8.8in}]{geometry}
\usepackage[utf8]{inputenc}

\usepackage{graphicx}
\usepackage{caption}
\usepackage{subcaption}
\usepackage[colorlinks=true,linkcolor=black,anchorcolor=black,citecolor=black,filecolor=black,menucolor=black,runcolor=black,urlcolor=black]{hyperref}

\usepackage{abstract}
\usepackage{authblk}

\usepackage{gensymb}

\newcommand{\supplementarysection}{%
  \setcounter{figure}{0}% Reset figure counter
  \let\oldthefigure\thefigure% Capture figure numbering scheme
  \renewcommand{\thefigure}{SI-\oldthefigure}% Prefix figure number with S
  \section*{Supplementary Information}% Set supplementary section
  \let\oldchapter\chapter% Copy \chapter into \oldchapter
}

\usepackage[citestyle=ieee,firstinits=true,backend=biber,maxnames=1,minnames=1]{biblatex}
\renewbibmacro{in:}{}

\title{High-power and narrow-linewidth laser on thin-film lithium niobate enabled by photonic wire bonding}
 
\author[1,2,*]{Cornelis A.A. Franken}
\author[1,*]{Rebecca Cheng}
\author[1]{Keith Powell}
\author[1]{Georgios Kyriazidis}
\author[3]{Victoria Rosborough}
\author[3]{Juergen Musolf}
\author[1,4]{Maximilian Shah}
\author[1,5]{David R. Barton III}
\author[1]{Gage Hills}
\author[3]{Leif Johansson}
\author[2]{Klaus-J. Boller}
\author[1,6]{Marko Lon\v{c}ar}

\affil[1]{John A. Paulson School of Engineering and Applied Sciences, Harvard University, Cambridge, MA 02138, USA}
\affil[2]{Laser Physics \& Nonlinear Optics, Department of Science and Technology, MESA+ Institute of Nanotechnology, University of Twente, Enschede, NL}
\affil[3]{Freedom Photonics, 41 Aero Camino, Goleta CA 93117, USA}
\affil[4]{Williams College, Williamstown, MA 01267, USA}
\affil[5]{Department of Materials Science and Engineering, Northwestern University, Evanston, IL 60208, USA}
\affil[6]{Corresponding author: loncar@seas.harvard.edu}
\affil[*]{These authors contributed equally to this work.}

\date{July 5, 2024}

\begin{document}
\maketitle

\vspace{-0.25cm}
\begin{abstract}
\noindent Thin-film lithium niobate (TFLN) has emerged as a promising platform for the realization of high-performance chip-scale optical systems, spanning a range of applications from optical communications to microwave photonics. Such applications rely on the integration of multiple components onto a single platform. However, while many of these components have already been demonstrated on the TFLN platform, to date, a major bottleneck of the platform is the existence of a tunable, high-power, and narrow-linewidth on-chip laser. Here, we address this problem using photonic wire bonding to integrate optical amplifiers with a thin-film lithium niobate feedback circuit, and demonstrate an extended cavity diode laser yielding high on-chip power of 78 mW, side mode suppression larger than 60 dB and wide wavelength tunability over 43 nm. The laser frequency stability over short timescales shows an ultra-narrow intrinsic linewidth of 550 Hz. Long-term recordings indicate a high passive stability of the photonic wire bonded laser with 58 hours of mode-hop-free operation, with a trend in the frequency drift of only 4.4 MHz/h. This work verifies photonic wire bonding as a viable integration solution for high performance on-chip lasers, opening the path to system level upscaling and Watt-level output powers.
\end{abstract}
\vspace{0.25cm}

\subsection*{Introduction}
\noindent Integrated photonics shows promise for the realization of low-cost, energy-efficient, and scalable solutions for numerous applications across communications, computation, and sensing. In the past decade, the thin-film (TF) lithium niobate (LN) platform has positioned itself as a promising alternative to silicon on insulator, boasting many material properties that other platforms lack: its large electro-optic coefficient has made it an excellent host for efficient high-speed, wide-bandwidth modulators; its large second order nonlinearity and ability to be periodically poled allows for efficient nonlinear three-wave mixing processes; and its wide transparency window allows for low-loss propagation over a wide range of wavelengths \cite{zhu2021integrated}. Recently, a slew of TFLN components have been demonstrated, including high-performance electro-optic modulators \cite{wang2018integrated,xu2020high}, broadband supercontinuum \cite{okawachi2020chip,jankowski2020ultrabroadband} and Kerr comb \cite{he2019self,song2024octave} generation, nonlinear frequency conversion \cite{chen2024adapted,xin2024wavelength}, and more, confirming it as an excellent platform for large-scale and diverse photonic systems. Before fully-integrated systems on TFLN can be realized, a key element to drive on-chip systems remains an outstanding challenge: a tunable, narrow-linewidth and high-power laser that removes the need for an off-chip, bulk-optic source. 

\begin{figure}[t]
\centering
\includegraphics[width=110mm]{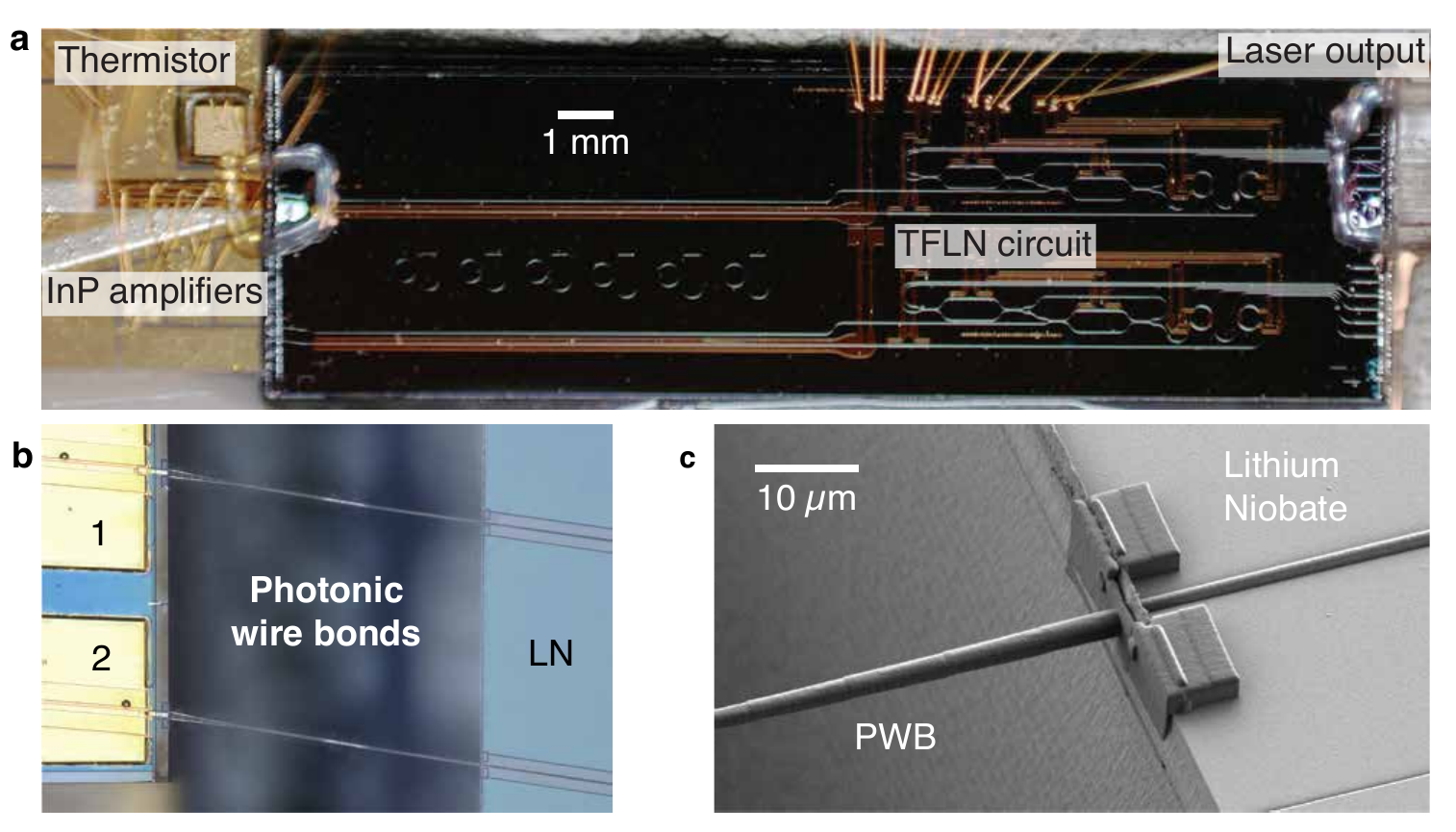}
\caption{\textbf{Photonic wire bonded TFLN-based extended cavity laser. }\textbf{a,} Photograph of the fully integrated laser realized using an InP chip with two amplifier waveguides and a TFLN feedback circuit, with photonic and electrical wire bonds. The temperature is monitored and stabilized using a thermistor and Peltier cooler. The laser output is collected using a fiber array. \textbf{b, }Microscope image of the InP-LN interface, where 1 and 2 denote the two amplifiers in the laser cavity, with photonic wire bonds and waveguides angled to mitigate Fresnel reflections. \textbf{c, }Scanning electron microscope image of a photonic wire bond. with anchoring structure, to a TFLN waveguide.}
\label{fig:ecl_fig1}
\end{figure}

As integrated photonic systems continue to scale up, co-integration of lasers and various TFLN devices would eliminate the large cumulative insertion losses from using bulk components, increasing signal-to-noise ratio and resulting in more compact, energy efficient, and higher performance on-chip systems \cite{Marpaung2019}. Due to the lack of a direct bandgap in lithium niobate, different laser integration approaches have been pursued. Aside from rare-earth doped lithium niobate lasers \cite{Zhou2022,Luo2023}, the majority of TFLN laser development has been focused on integration methods with semiconductor gain materials, including heterogeneous \cite{opdebeeck2021iii,morin2024coprocessed} and hybrid integration \cite{li2022integrated,shams2022electrically, ling2023self,han2024integrated}. However, significant improvement is still needed to achieve the combination of high performance with scalability. Devices using heterogeneous integration approaches require complex, wafer scale fabrication strategies and have yet to show high-power operation \cite{Zhang2023}. On the other hand, traditional hybrid strategies relying on precise, sub-wavelength alignment accuracy of components have shown promising performance, %using stage-coupled components 
but have yet to be effectively adopted at the wafer scale due to these stringent alignment requirements \cite{Porter2023}. While required performance metrics are application-specific, as a benchmark for laser performance we use current commercial bulk-optic systems, which offer in the order of 100 mW fiber-coupled output power, with gain-wide tunability and linewidths in the sub 100-kHz range. Besides the performance metric, here ``scalability'' refers not only to a fabrication and integration process that allows for high-volume production, but also to the capability of straightforward integration of many separate components, from optical amplifiers and modulators to fiber arrays. 

An emerging integration technology that could address scaling issues without compromising performance is photonic wire bonding (PWB), which enables three-dimensional and low-loss optical connections between different photonic chips via polymer waveguides \cite{lindenmann2012photonic}. Importantly, the PWB formation process not only can take into account the relative position of different facets, but also can be adapted to match different optical mode profiles. Therefore, it can overcome large misalignment and mode mismatch between a whole range of components within a system, including but not limited to amplifiers, waveguides, and fibers. Previous work has already heralded the strength of PWB combined with silicon photonics, showing its applicability for high confinement platforms \cite{billah2018hybrid, xu2021hybrid,chowdhury2024chip}. However, the TFLN platform, with its many functionalities, enables PWB to tailor to a much wider range of applications.

Here, we present a photonic wire bonded, extended cavity laser that integrates indium phosphide (InP) amplifiers with a lithium niobate feedback circuit (Fig. \ref{fig:ecl_fig1}a). As an initial demonstration of the flexibility of the photonic wire bonding integration process, we realize upscaling in performance and integration through PWB of a second amplifier waveguide to one cavity (Fig. \ref{fig:ecl_fig1}b), allowing for power scaling in such lasers with possibility of adding additional amplifiers as needed.

\subsection*{Design, fabrication and integration}
\begin{figure}[t]
\centering
\includegraphics[width=110mm]{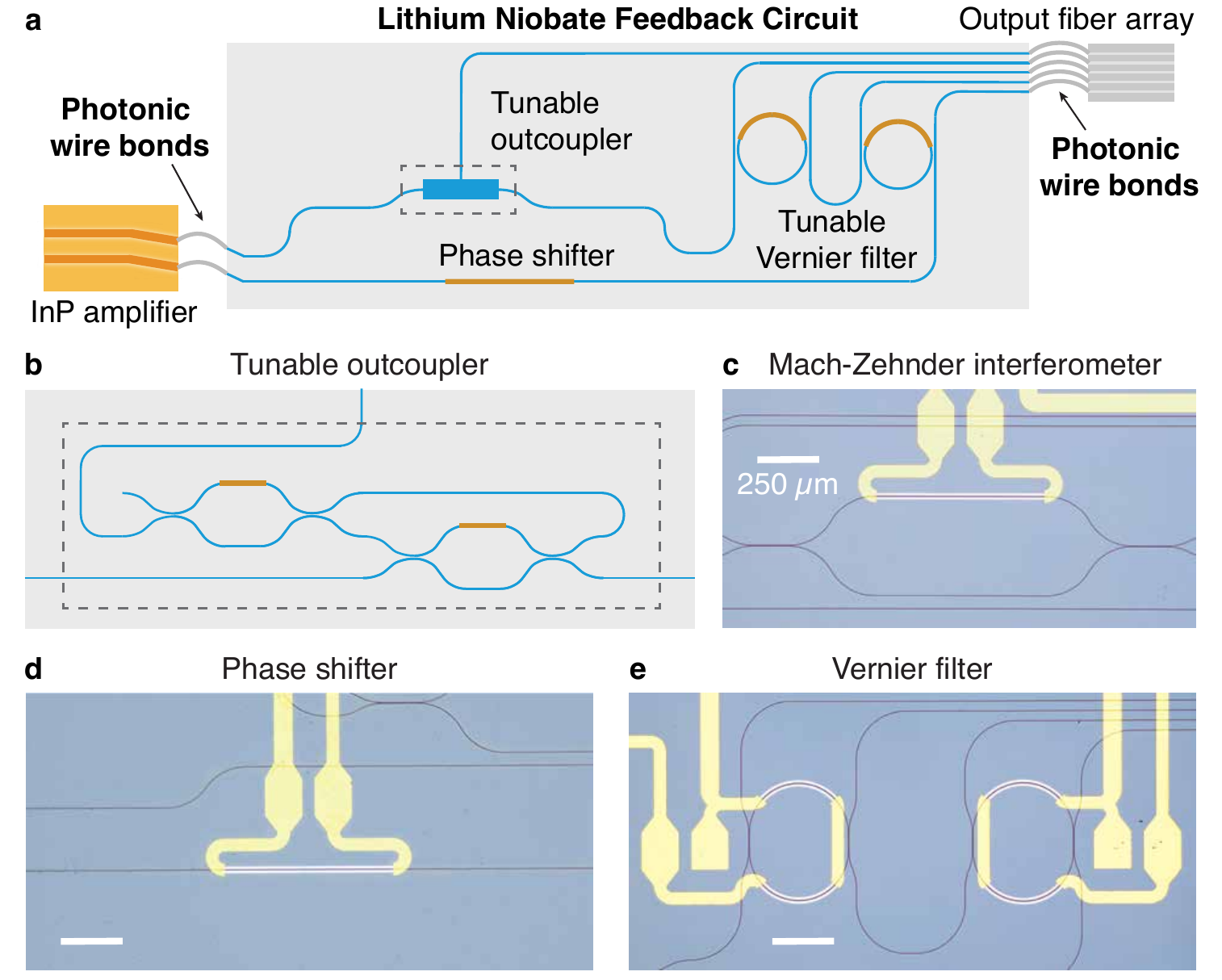}
\caption{\textbf{Lithium niobate extended laser cavity. }\textbf{a,} Schematic showing the design of the laser, which comprises of two InP amplifier waveguides, an LN feedback circuit and a fiber array, with all components integrated by photonic wire bonding. The main feature of the feedback circuit is the Vernier filter, consisting of two coupled micro-ring resonators, which provides frequency selective feedback and cavity extension to the laser. The effective length of laser cavity, center frequency of the filter, and outcoupling is finely adjusted using thermo-optic tuning of the phase shifter, Vernier filter and tunable outcoupler, respectively. \textbf{b, } Schematic of the tunable outcoupler, realized through the use of two tunable Mach-Zehnder interferometers (MZI). \textbf{c,} \textbf{d,} and \textbf{e,} Microscope images of one of the MZIs, the phase shifter, and Vernier filter, respectively.}
\label{fig:ecl_fig2}
\end{figure}

A major advantage of hybrid integration is the separate fabrication of optical amplifiers and photonic feedback circuits. This benefit remains for our laser, allowing the amplifiers to be fabricated in a dedicated semiconductor foundry for uncompromised performance. Likewise, our lithium niobate fabrication process needs no adaptation for the laser integration process. After fabrication of the individual components, the integration process starts by coarsely placing and aligning the individual amplifier, TFLN, and fiber array on a common submount. The tool applying the photonic wire bonds (Vanguard Sonata 1000) offers a large write field, allowing for a wide tolerance of the optical interface alignment of approximately 320 x 325 x 250 $\mu$m for the $x$-, $y$-, and $z$- direction, respectively (PWB is written along the $x$-direction, and $z$ is height). The position and angle of the optical components are detected using a confocal microscope and machine vision algorithm. The topside of the InP is partially obscured due to a metal layer used for applying current to the amplifier, reducing the detection accuracy when compared to the TFLN waveguide and fiber core. Therefore, photonic wire bond alignment on the amplifier side typically requires a refined calibration of the photonic wire bond position using experimentally obtained transmission data. Such a calibration is required only once for a set amplifier material stack and amplifier waveguide geometry. Using the location of the facets and user input on the mode profiles at each interface, the PWB structure is consequently written in a negative photoresist using two-photon polymerization with a pulsed near-infrared laser. The PWB structure is routed to maximize the transmission, by ensuring a large bend radius ($\mathrm{>}$ 60 $\mu$m) and using sufficiently long tapers ($\mathrm{>}$ 100 $\mu$m per taper). After development of the resist, for endured stability and increased transmission, all PWB interfaces are cladded with a UV-cured epoxy. A scanning electron microscope (SEM) image of a photonic wire bond to a TFLN spot size converter \cite{he2019low} before cladding is shown in Fig. \ref{fig:ecl_fig1}c. The single-pass PWB loss between amplifier and LN is measured at 4 dB, while the LN to fiber PWB loss is 3 dB. To our knowledge, these are the lowest photonic wire bonding losses reported so far for the thin-film lithium niobate platform interfacing with a fiber or semiconductor amplifier.

The TFLN waveguide circuit provides the cavity extension and frequency selection in the laser using a Vernier filter consisting of two tunable, micro-ring resonators \cite{boller2020hybrid}. A full schematic of the dual-gain laser design is shown in Fig. \ref{fig:ecl_fig2}a, which depicts the Vernier filter with two sequentially coupled micro-ring resonators with a difference in free spectral range (FSR) of approximately 2 GHz. The Vernier filter allows for coarse tuning of the wavelength while, the effective length of the laser cavity, hence cavity resonance, can be fine-tuned using the thermo-optic phase shifter. Optimizing the output coupling of the laser allows for maximum extraction of optical power at a set current and Vernier filter setting, this is achieved using a tunable outcoupler. In our design, this outcoupler is realized through two cascaded tunable Mach-Zehnder interferometers (MZI), giving full control to the splitting of feedback to the amplifier and power to the output waveguide (Fig. \ref{fig:ecl_fig2}b).  The rings, couplers, and cavity length are actively tuned using Ti-Pt thermo-optic heaters (Fig. \ref{fig:ecl_fig2}c,d,e). Separate characterization on a Mach-Zehnder test structure shows a phase shifting efficiency of about 125 mW/$\pi$ and full control over the power splitting between 0 and 100\%. Heaters at the ring resonators allow for shifting the resonance frequency over the full FSR (measured at around 113 GHz), with an efficiency of about 150 mW/$\pi$. Using measured transmission spectra of ring resonators fabricated on the same chip, we infer a low propagation loss in the TFLN circuit of 5.8 dB/m. The low loss is mainly attributed to the high confinement of the optical mode and low sidewall roughness. After photonic wire bonding all optical components, the packaging is completed using gold wire bonds for the electrical connections and a Peltier element with on-chip thermistor for temperature control.

\begin{figure}[t!]
\centering
\includegraphics[width=110mm]{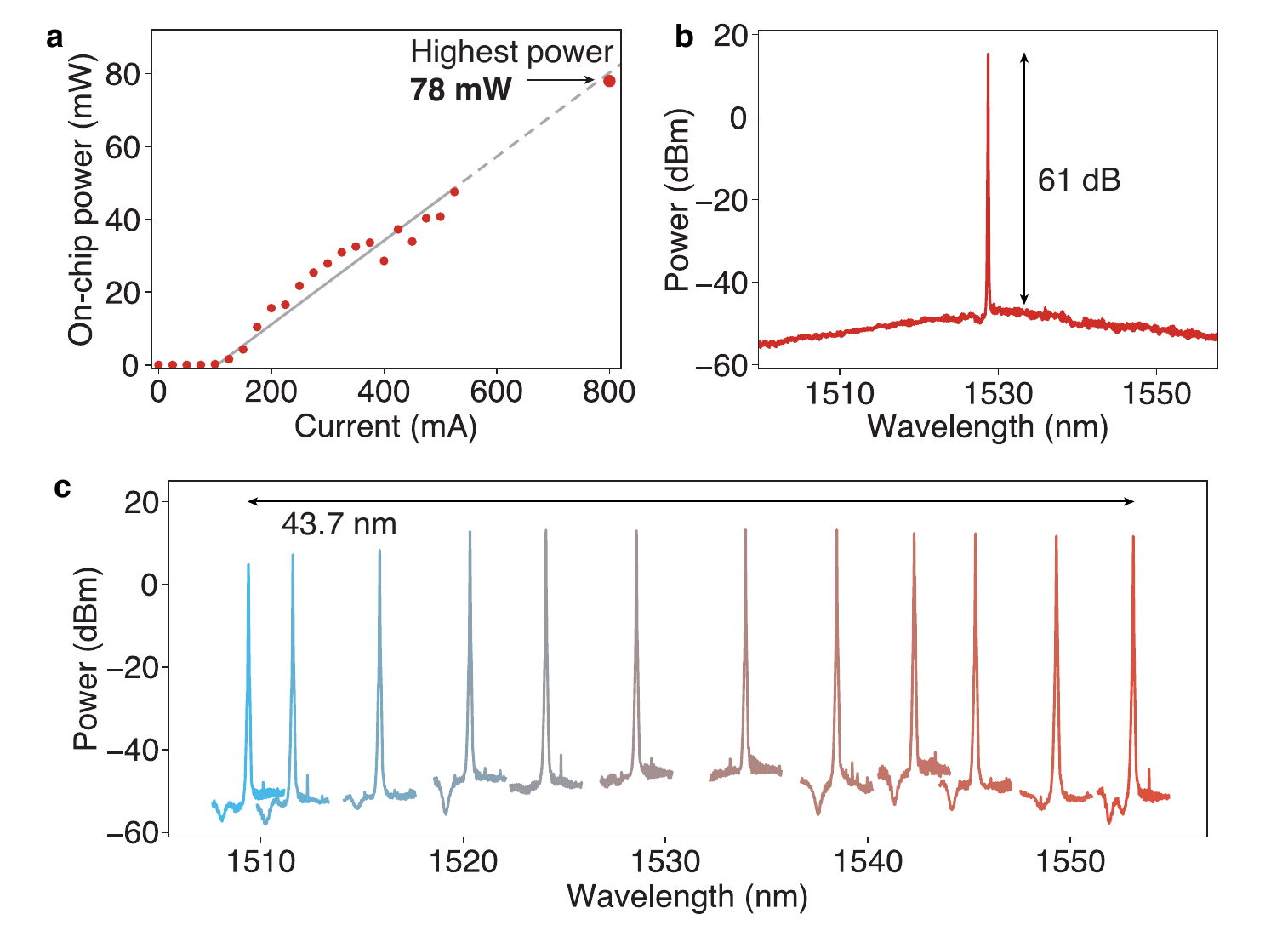}
\caption{\textbf{Laser power, side mode suppression, and wavelength tuning.} \textbf{a,} On-chip optical power as a function of current applied to each amplifier. With a laser threshold of 100 mA, the laser shows a linear power increase as a function of current. Residual variations to the linear trend are caused by longitudinal mode hops. The maximum power is obtained by setting the amplifiers at different currents, 200 mA and 800 mA, respectively. \textbf{b, }Optical spectrum showing single-frequency operation with 61 dB side mode suppression. \textbf{c, } Superimposed optical spectra obtained by varying the heater setting for one of the ring resonators in the Vernier filter, showing a wide wavelength tunability of 43.7 nm.}
\label{fig:ecl_fig3}
\end{figure}

\subsection*{Experimental results}
With the laser in operation, we find a laser threshold at 100 mA and an approximately linear increase of output power with current (Fig. \ref{fig:ecl_fig3}a). At each current, the Vernier filter central wavelength is optimized together with the cavity phase section to obtain single-mode oscillation with maximum output power and side mode suppression. To achieve the highest power measurement, different currents were applied to the amplifiers, 200 mA and 800 mA respectively, at which a high, on-chip output power of 78 mW was achieved. Optical spectra (Yokogawa AQ6370) confirm that, at all currents reported here, the laser oscillates single longitudinal mode with a high side mode suppression ratio up to 61 dB (Fig. \ref{fig:ecl_fig3}b). The laser wavelength can be tuned coarsely by varying the effective length of one the ring resonators in the Vernier filter using a thermo-optic heater. Here, the laser is widely tuned in wavelength around 1530 nm over 43.7 nm, which is beyond the 3-dB gain bandwidth of both amplifiers. Superimposed spectra collected at different settings for the Vernier filter central wavelength is given in Fig. \ref{fig:ecl_fig3}. A high optical power, 15 mW on average, and single mode operation with $\mathrm{>}$52.7 dB side mode suppression across the entire tuning range is measured.

\begin{figure}[t!]
\centering
\includegraphics[width=110mm]{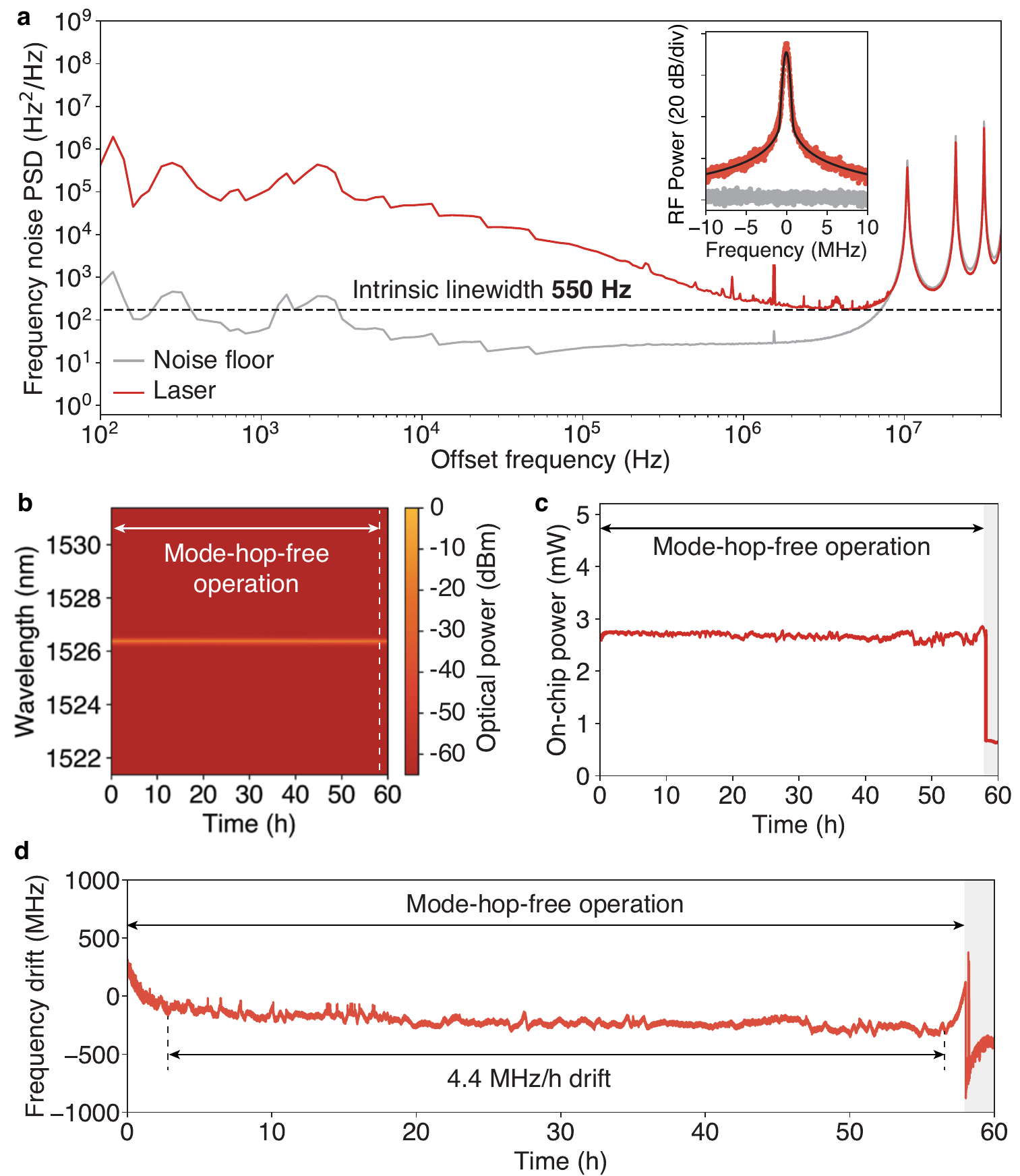}
\caption{\textbf{Laser frequency stability over short and long timescales}. \textbf{a, }Frequency noise power spectral density acquired using delayed self-heterodyne measurement, showing an ultra-narrow 550 Hz intrinsic linewidth. Inset: beat note recorded in the frequency domain with a Voigt fit yielded a similar intrinsic linewidth. \textbf{b,} As part of a multi-day long term recording, the optical spectrum was recorded. Here, a cumulative spectral plot shows the laser spectrum during the 58 hours of mode-hop-free operation with a sustained, high side mode suppression ratio of 55.5 $\mathrm{\pm}$ 0.4 dB. \textbf{c,} On-chip optical power recording during the long measurement. We show the region of mode-hop-free operation and denote the observed mode hop in grey. \textbf{d,} Long term frequency drift of the laser collected using a high-resolution wavelength meter. The laser exhibits mode-hop-free operation for 58 hours. After initialization and before the mode hop, we observe a trend in the frequency drift of 4.4 MHz/h. The observed mode hop is highlighted in grey. }
\label{fig:ecl_fig4}
\end{figure}

Equally important to output power and wavelength tunability is the passive long and short-term frequency stability of the laser to enable low-noise devices, necessary for example in microwave photonics applications \cite{Marpaung2019}. Here, the short term frequency stability can be characterized by measuring the intrinsic linewidth of the laser \cite{Schawlow1958a,boller2020hybrid}. Due to the long, extended cavity of our laser design we expect a low intrinsic linewidth, calculated to be in the order of 100 Hz. However, the resolution of the optical spectrum analyzer (20 pm or 2.5 GHz) used for the measurements shown in Fig. \ref{fig:ecl_fig3} is not sufficient to resolve the linewidth of the laser. To resolve such an ultra-low intrinsic linewidth a delayed self-heterodyne set-up is employed \cite{VanRees2023} with a 19.6 m length delay. The measured frequency noise power spectral density, computed from five 100-ms long oscilloscope (Keysight UXR1102B) traces of the heterodyne beat note \cite{Yuan2022}, is shown in Fig. \ref{fig:ecl_fig4}a. For this measurement, the amplifiers of the laser are set to respectively 225 and 150 mA and the laser produces an on-chip power of 21.3 mW at a wavelength of 1527.8 nm. The frequency noise flattens at the quantum-limited noise around 2 MHz, at a level that corresponds to an ultra-narrow intrinsic linewidth of 550 Hz. A similar linewidth, within the experimental error, is measured by recording the beat note in the frequency domain using a long delay of 20 km and performing a Voigt fit to the line shape. We note that for both cases, we are not limited by the noise floor of our measurement. 

In contrast to mechanical, acoustic and thermal alignment drifts of bulk-optic extended cavity lasers, integrated photonic circuits typically show excellent passive and long-term stability, an essential benefit for the scaling to large, fully-integrated systems. Previous demonstrations of chip-based external cavity diode lasers on silicon nitride have realized stable mode-hop-free operation of 5.7 hours for stage-coupled lasers \cite{corato2023widely} and up to 120 hours for actively stabilized, integrated lasers \cite{VanRees2023}. To investigate the passive, long-term frequency stability of the laser without any active stabilization, we operate and monitor its performance continuously over multiple days. The current on both amplifiers is set just above threshold at 125 mA current to minimize thermal fluctuations, corresponding to an initial on-chip output power of 2.6 mW. After the thermo-optic heaters are set, the laser output is monitored continuously using an optical spectrum analyzer (Yokogawa AQ6370), power meter (Thorlabs S155C), and a high resolution wavelength meter (High Finesse WS6-200 IR, resolution of 5 MHz) to assess the spectral characteristics, optical power, and frequency drift, respectively. Following initial start up and when thermal equilibrium is reached, the laser maintains stable and mode-hop-free over a 58-hour period (Fig. \ref{fig:ecl_fig4}b, c ,d). The laser emission spectrum over time is shown in a cumulative spectral plot in Fig. \ref{fig:ecl_fig4}b, with a sustained, high side mode suppression ratio of 55.5 $\mathrm{\pm}$ 0.4 dB during the mode-hop-free operation. The on-chip optical power over time is given in Fig. \ref{fig:ecl_fig4}c, with less than $\pm0.2$ mW variation. Finally, Fig. \ref{fig:ecl_fig4}d shows the frequency drift of the laser extracted from the wavemeter measurement. We note excellent passive frequency stability, with a trend in the frequency drift of 4.4 MHz/h.

\subsection*{Discussion and conclusion}
In conclusion, we demonstrate a high-power, passively stable, and ultra-narrow linewidth extended cavity laser integrated with photonic wire bonding. The maximum on-chip output power exceeds existing TFLN extended cavity diode lasers by 1-2 orders of magnitude \cite{li2022integrated, opdebeeck2021iii}, due to the power scaling of the dual amplifier design enabled by photonic wire bonding. To our knowledge, our on-chip power exceeds all single-gain ECLs across all photonic platforms, with performance comparable to or approaching that of other dual-gain ECLs demonstrated on the more mature silicon and lower-loss silicon nitride platforms \cite{boller2020hybrid, kobayashi2015silicon, zhao2021hybrid}. Currently, our laser shows the lowest intrinsic linewidth of any on-chip laser source demonstrated using TFLN. We attribute the narrow linewidth mainly to the low-loss and long extended cavity, with an effective length of about 24 cm. The highly frequency selective Vernier design allows the laser wavelength to be tuned over twice the range of what has been reported before \cite{li2022integrated}. As far as we know, this work is a first demonstration of long-term frequency stability for TFLN-based extended cavity diode lasers. Our lithium niobate laser also shows comparable or improved passive stability to similar Vernier-based lasers demonstrated using silicon nitride \cite{VanRees2023, Gonzalez-Guerrero2022}.

Although all of the measurements presented here show single frequency operation with high side mode suppression, we found that at higher currents ($\mathrm{>}$ 525 mA) the laser is more likely to oscillate multi-mode, similar to what has been observed in other recent work \cite{Xue2024}. The reason for the decreased stability regime at higher powers is likely due to the coupling loss between the InP and TFLN waveguides, which could be improved upon in future iterations. Furthermore, it became apparent that the laser performance had degraded with time, seen as a lower maximum power for measurements taken after the initial 78 mW measurement. We attribute this to gradual moisture diffusion into the top oxide cladding leading to increased absorption and optical loss. Currently, high-quality LPCVD oxides are difficult to integrate with the lithium niobate fabrication process due to the high temperatures required for deposition. As a consequence, we resort to a lower temperature PECVD $\mathrm{SiO_2}$. To suppress the moisture diffusion into the top oxide, an aluminum oxide sealing layer, deposited by ALD, was used to slow the degradation process down, though it does not eliminate it fully. To assess the quality of the oxide cladding and the efficacy of the sealing layer, optical $Q$ measurements on the TFLN chip are used to evaluate loss. Immediately after processing, the inclusion of the PECVD oxide cladding leads to a 20\% reduction in the optical $Q$ compared to the uncladded case, pointing to reduced optical quality of our current oxide. Moreover, with the ALD layer, the optical $Q$ degrades by 15\% over two weeks of aging in environment compared to $\mathrm{>}$50\% with no sealing layer. A long hot plate bake ($\mathrm{>}$85$^\circ$C) before integration can recover any degradation of the $Q$, but is currently unfeasible for the fully integrated device. We believe that higher optical powers could be achieved reproducibly with the implementation of higher density oxides that are compatible with the lithium niobate fabrication process.

Looking forward to higher power applications, the slope efficiency and maximum laser power can be improved significantly by increasing the PWB transmission and reducing the LN propagation loss to $\mathrm{<}$3 dB/m \cite{zhang2017monolithic,shams2022reduced,zhu2024twenty}, which would lower the intra-cavity loss. We expect that in future demonstrations, PWB loss can be improved through further optimization of the bond alignment and improvement of the facet quality. Moreover, losses could be reduced further with the improvement to the TFLN coupler design. In further experiments, the PWB showed no degradation with up to 1 W of optical power in the bond, verifying that high intra-cavity powers could be supported. Furthermore, photonic wire bonding to tapered amplifiers could open the path towards optical powers in the Watt-level regime. 

In addition to completing the TFLN platform with a tunable, narrow-linewidth and high-power laser, our work verifies photonic wire bonding as a scalable solution for performance and integration. The power scaling through photonic wire bonding can be used to enable different nonlinear processes, while a combination with fast modulation can be useful for microwave photonic applications \cite{Marpaung2019}. Future work involves integration with optical isolators \cite{yu2023integrated} for a fully protected laser cavity, as well as other functionalities of the TFLN platform, such as high performance electro-optic modulators \cite{wang2018integrated,kharel2021breaking}, electro-optic frequency comb sources \cite{hu2022high,yu2022integrated}, and high-speed on-chip detectors \cite{guo2022high}.
\subsection*{Acknowledgments} The authors thank Jeremiah Jacobson for the photograph in Fig. \ref{fig:ecl_fig1}a, Donald Witt for the image in Fig. \ref{fig:ecl_fig1}c, Xinrui Zhu and Yaowen Hu for providing a characterization chip for the LN-fiber PWB loss, Allen Chu and Henry Garrett for initial PWB characterization, and Yunxiang Song, Lisa V. Winkler, Let\'{i}cia Magalh\~{a}es, and Amirhassan Shams-Ansari for support during the project. Lithium niobate fabrication was performed in the Center for Nanoscale Systems at Harvard University.
\subsection*{Funding}  Defense Advanced Research Projects Agency (DARPA) HR0011-20-C-0137, Office of Naval Research (ONR) N00014-22-C-1041, National Science Foundation (NSF) EEC-1941583 and OMA-2137723.
\subsection*{Disclosures} V.R., J.M., and L.J. are involved in developing III-V technologies at Freedom Photonics. R.C. and M.L. are involved in developing lithium niobate technologies at HyperLight Corporation.
\subsection*{Data availability} Data available upon reasonable request.
\subsection*{Disclaimer} The views, opinions and/or findings expressed are those of the authors and should not be interpreted as representing the official views or policies of the Department of Defense or the U.S. Government.

% Bibliography
\vspace{-0.25cm}

\printbibliography

\end{document}